\documentclass[prd,twocolumn,floats,superscriptaddress,eqsecnum,floatfix,
showpacs]{revtex4-1}

\pdfoutput=1
\usepackage{amssymb,amsmath}
\usepackage{epsfig}
\usepackage{color}
\usepackage{datetime}

\newcommand{\be}{\begin{equation}}
\newcommand{\ee}{\end{equation}}
\newcommand{\ba}{\begin{eqnarray}}
\newcommand{\ea}{\end{eqnarray}}
\newcommand{\beg}{\begin{gather*}}
\newcommand{\eng}{\end{gather*}}

\newcommand{\hh}{,\hspace{0.5cm}}
\newcommand{\hhh}{,\hspace{0.2cm}}

\newcommand{\eq}[1]{(\ref{#1})}

\newcommand{\lap}{\bigtriangleup}

\newcommand{\n}[1]{\label{#1}}

\newcommand{\ins}[1]{{\mbox{\tiny #1}}}
\newcommand{\ind}[1]{{\mbox{\scriptsize #1}}}
\newcommand{\inds}[1]{{\scriptscriptstyle #1}}

\begin{document}

\title{Head-on collision of ultrarelativistic particles in ghost-free theories of 
gravity}

\author{Valeri P. Frolov}
\email{vfrolov@ualberta.ca}
\affiliation{Theoretical Physics Institute, Department of Physics\\
University of Alberta, Edmonton, Alberta, Canada T6G 2E1}

\author{Andrei Zelnikov}
\email{zelnikov@ualberta.ca}
\affiliation{Theoretical Physics Institute, Department of Physics\\
University of Alberta, Edmonton, Alberta, Canada T6G 2E1}

\begin{abstract}
We study linearized equations of a ghost-free gravity in four- and higher-dimensional 
spacetimes.
We consider versions of such a theory where the nonlocal modification of the $\Box$
operator has the form $\Box \exp[(-\Box/\mu^2)^N]$, where $N=1$ or $N=2n$. We first
obtain the Newtonian gravitational potential for a point mass for such models and
demonstrate that it is finite and regular in any number of spatial dimensions $d\ge
3$. The second result of the paper is calculation of the gravitational field of an
ultrarelativistic particle in such theories. And finally, we study a head-on
collision of two ultrarelativistic particles. We formulated conditions of the
apparent horizon formation and showed that there exists a mass gap for 
mini-black-hole production in the ghost-free theory of gravity. In the case when the 
center-of-mass energy is sufficient for the formation of the apparent horizon, the 
latter has
two branches, the outer and the inner ones. When the energy increases the outer
horizon tends to the Schwarzschild-Tangherlini limit, while the inner horizon becomes
closer to $r=0$.
\end{abstract}


\maketitle

\section{Introduction}

Singularities are inherent properties of general relativity. It is generally
believed that the Einstein-Hilbert action should be modified in spacetime domains 
where the curvature becomes large. Such a modification is required, for example, when 
one includes in the theory quantum corrections, connected with particle creation and 
vacuum polarization effects. At a more fundamental level, the modification of the
gravity equation might be required if the gravity is described as an emergent
phenomenon. In such a case the Einstein equations are nothing but the low energy
limit of the corresponding more fundamental background theory. The string theory is a
well-known example. It is convenient to introduce two (generally different) energy
scale parameters $\mu$ and $\tilde{\mu}$. The corresponding length scales are
$\lambda=\mu^{-1}$ and $\tilde{\lambda}=\tilde{\mu}^{-1}$. We assume that when the
spacetime curvature ${\cal R}$ is much less than $\lambda^{-2}$, the corrections to
the Einstein equations are small. These corrections become comparable with other
terms of the Einstein equations at ${\cal R}\sim \lambda^{-2}$, and for higher values
of the curvature they play an important role. We assume that one can use the 
classical
metric $g_{\mu\nu}$ for the description of the gravitational field. For example, one
can understand it as a quantum average of some metric operator, $g_{\mu\nu}=\langle
\hat{g}_{\mu\nu}\rangle$. This means that the quantum gravity effects, and in
particular fluctuation of the metric, are small. In other words, one can use the
effective action approach to study spacetime properties in this domain. The second
parameter, $\tilde{\lambda}$, defines the scale when effective action description
breaks down and the quantum nature of the gravitational field becomes important.

In studies of the singularity problem in modified gravity it is usually assumed that
$\tilde{\mu}\gg\mu$. In the present paper we also use this assumption and discuss
some aspects of the singularity problem in the framework of the classical modified
gravity equations.

There exist a wide class of the modified theories of gravity proposed to solve
fundamental problems of black holes and cosmology. We consider a special class of 
such
theories, namely theories with higher derivatives. Important features of such
theories can be clarified already in a simple approximation when the gravitational
field is weak and can be described as the perturbation on the flat spacetime
background. Such an analysis was performed by Stelle \cite{Stelle:1977ry}. In
particular, he demonstrated that the Newtonian gravitational potential of a point
mass located at $\vec{r}=0$ can be made finite at this point, if the higher
derivative terms are included in the gravity equations. Detailed analyses of this
problem can be found in recent papers \cite{Modesto:2014eta,Asorey:1996hz}.

However, the higher derivative gravity, as well as any theory with higher
derivatives, has a fundamental problem. In a general case the propagator of such a
theory contains two or more poles, and, as a result, it almost always contains
ghostlike excitations (see, e.g., \cite{Stelle:1977ry} and
\cite{Biswas:2005qr},\cite{Barnaby:2007ve}). Presence of the excitations
with negative energy results in an instability of the theory and the possibility of 
an
empty space decay.
This is a special case of a very general
phenomenon known as Ostrogradsky instability \cite{Ostrogradsky:1850}
(see discussion in \cite{Barnaby:2007ve}).

 In the higher-derivative theory a standard box operator $\Box$, which enters the
field equations is changed to the operator $P(\Box)\Box$, where $P(z)$ is a
polynomial. The poles of $P^{-1}(z)$ correspond to additional degrees of freedom.
However, there exists an
interesting option of theories where $P^{-1}(z)$ is an entire function of $z$ and 
hence
it does not have poles in the complex plane. Such a modification of the gravitational
equations is called ghost-free (GF) gravity (see, e.g.,\cite{Tomboulis:1997gg,
Biswas:2011ar, Modesto:2011kw, Modesto:2012ys, Biswas:2013cha, Biswas:2013kla,
Modesto:2014lga,Tomboulis:2015gfa,Tomboulis:2015esa} and references therein).
GF gravity contains an infinite number of derivatives and, hence, it is nonlocal.
Theories
of this type were considered a long time ago
(see, e.g.,
\cite{efimov1967,Efimov:1972wj,Efimov:1976nu,Efimov:18,
Efimov:19}).
They appear naturally also in the context of noncommutative geometry deformation of the
Einstein gravity \cite{Nicolini:2005vd,Spallucci:2006zj} (see a review \cite{Nicolini:2005zi} and references therein).
The initial value problem in nonlocal theories was studied in
\cite{Barnaby:2007ve,Barnaby:2010kx}.The
application of the ghost-free theory of gravity to the
problem of singularities in cosmology and black holes can be found in
\cite{Biswas:2010zk, Modesto:2010uh, Hossenfelder:2009fc, Calcagni:2013vra,
Zhang:2014bea, Conroy:2015wfa,Li:2015bqa}. Static and dynamical solutions of the
linearized equations of the ghost-free gravity in four and higher dimensions were
studied in \cite{Frolov:2015bta,Frolov:2015bia}.
Recently the consequences of the ghost-free modifications
of higher-dimensional gravity
on the entropy of black holes and on cosmological models have been studied \cite{Conroy:2015nva}.

In this paper we continue study of the linearized equations of the GF gravity. In
Secs. \ref{sec2}-\ref{sec4} we study solutions for a static gravitational field in
the Newtonian
approximation in different models of the GF gravity. Namely, we consider a class of
the $\mathrm{GF_\inds{N}}$ theories of gravity with $P(\Box)=\exp[(-\Box/\mu^2)^N]$. 
A static
solution of the linearized equations for $N=1$ in four-dimensional spacetime was
found in \cite{Biswas:2011ar,Modesto:2010uh} (see also \cite{Frolov:2015bia}). In
this paper we generalize this result to the higher-dimensional case and obtain new
solutions for $\mathrm{GF_{2n}}$ theories in the spacetime with an arbitrary 
number $d$ of
spatial dimensions. In Sec. \ref{sec5} we used these results to obtain a solution
of the GF
gravity describing a gravitational field of an ultrarelativistic particle. We
succeeded to find a generalization of the famous Aichelburg-Sexl solution
\cite{Aichelburg:1970dh} to the GF gravity in an arbitrary number of dimensions.
In 
Sec. \ref{sec6} we used the obtained solutions to study the apparent horizon
formation
in head-on collision of two ultrarelativistic particles. This problem for the
general
theory of relativity in four dimensions was first solved by Penrose \cite{Penrose}.
Later,  this result was generalized for a collision with a nonzero impact parameter 
in
four and higher dimensions 
\cite{Eardley:2002re,Yoshino:2002tx,Yoshino:2005hi,Constantinou:2013tia}. In
the present paper we show that in the GF gravity a similar process has two important
new features: (i) the apparent horizon is not formed if the center-of-mass energy of
the particles, $E$, is smaller than some critical value $E_{crit}$, which depends
on the scale
parameter $\mu$, the type of the theory, and the number of spacetime
dimensions; (ii) if the energy is larger than $E_{crit}$
the apparent horizon besides the usual outer part always has another inner branch. We
discuss the obtained results in the last section.

In the present paper we use units in which $\hbar=c=1$ and sign conventions adopted
in the book \cite{Misner:1974qy}.

\section{Newtonian limit of higher-dimensional higher-derivative
equations}\label{sec2}

Let us consider a static gravitational field perturbation on a flat background and
write the corresponding metric in the form
\be\begin{split}\label{WFM}
ds^2&=-(1+2\varphi)\,dt^2+(1-2\psi+2\varphi)\,d\ell^2,\\
d\ell^2&=\delta_{ik}\,dx^i dx^k\hh x^i=(x^1,\dots,x^{d}) .
\end{split}\ee
Here and later we denote by $d=D-1$ a number of spatial dimensions.
We also have
\begin{gather*}
h_{00}=-2\varphi\hh h_{ij}=-2(\psi-\varphi)\delta_{ij},\\
h=2[(d+1)\varphi-d\psi]\, .
\end{gather*}
By substituting these expressions into the gravity equations \eq{eqh} one gets
\begin{gather*}
a(\lap)\lap\psi=\kappa_d(\tau_{00}+{1\over d-1}\delta^{ij}\tau_{ij}),\\
[a(\lap)-dc(\lap)]\lap\varphi+(d-1) c(\lap)\lap\psi=\kappa_d\tau_{00}.
\end{gather*}
Here $\kappa_d=8\pi G^{(D)}$ and $D=d+1$ is the total number of spacetime dimensions.
In the Newtonian approximation $\delta^{ij}\tau_{ij}=0$ and the first of these
equations takes the form \be
a(\lap)\lap\psi=\kappa_d\tau_{00}\, .
\ee
For the gravity theory with $c=a$ the equations simplify and one obtains
\be
\psi={d-1\over d-2}\,\varphi\, ,
\ee
and the metric \eq{WFM} takes the form
\be
ds^2=-(1+2\varphi)\,dt^2+(1-{2\over d-2}\varphi)\,d\ell^2\, .
\ee
For a point mass $m$ the energy density has the form $\tau_{00}=m\delta^d(x)$. Then
for the Einstein gravity, where $a=c=1$ one has
\be\label{NGP}
\varphi=-{\kappa_d m\,\Gamma\left({d\over 2}\right)\over
2(d-1)\pi^{d/2}}\,{1\over r^{d-2}}\, .
\ee
In four dimensions $D=4$ $(d=3)$
\be
\varphi=-{\kappa_3 \over 8\pi}{m\over r}\, .
\ee

\section{Static solutions of linearized equations in ghost-free
gravity}\label{sec3}

\subsection{Ghost-free gravity}

The Newtonian potential \eq{NGP} is evidently singular at $r=0$. One can regularize
it and make it finite at $r=0$ by modifying the gravity equations in the ultraviolet
(UV) domain. For example, one may assume that $a(\Box)$ and $c(\Box)$ are
polynomials of the $\Box$ operator. If these functions obey the condition
$a(0)=c(0)=1$, the theory correctly reproduces the standard results of general
relativity in the infrared regime, that is in the domain where $r\to \infty$. In a
general case such a theory possesses ghosts. These ghosts are new degrees of freedom
which are connected with extra poles of the operators $a^{-1}$ and $c^{-1}$ which
give contributions to the propagator with a wrong (negative) sign. However, there
exists an option to use such functions $a^{-1}(z)$ and $c^{-1}(z)$ that are entire
functions of the complex $z$-variable which do not have poles. It happens, for
example, when $a(z)$ and $c(z)$ are of the form $\exp(P(z))$, where $P(z)$ is a
polynomial. A modified gravity which contains such regular formfactors is called
{\em ghost-free (GF) gravity}. In the present paper we focus on the special class of
the theories of GF gravity. Namely, we assume that
\be
a(\Box)=c(\Box)=\exp( (-\Box/\mu^2)^N)\, .
\ee
We denote such a theory $\mathrm{GF_\inds{N}}$. We
restrict ourselves by considering the cases $N=1$ and $N=2n$, which are of the most
interest for applications.

The exponent of the operator can be written in the form of a convergent series of
the powers of this operator. However, it is not a good idea to ``approximate" the
exponent by the polynomial which is obtained by keeping a finite number of terms in
this series. The inverse operator will have extra poles and the ghost will be
present for such truncation. That is why our first goal is to present these nonlocal
objects in the form of an integral transform which contains a well-defined kernel.


\subsection{Potential $\psi_d$ and Green functions in GF theories}

Consider the equation for the potential $\psi_d$ created by a point massive particle
placed at a point $x'$
\be\label{Fpsi}
\hat{F}\psi_d=\kappa_d  m\,\delta^d(x-x') ,
\ee
where the operator $\hat{F}$ is defined on the $d$-dimensional Euclidean space. It is
assumed to be a function of the Laplace operator \be
\hat{F}=\tilde{F}(-\lap)\hh \tilde{F}(\xi)=-\xi a(-\xi) .
\ee
The Euclidean Green function $D_d(x,x')$ of this operator is the solution of the 
problem
\be\label{FD}
\hat{F} D_d(x,x')=-\delta^d(x-x')
\ee
with vanishing boundary conditions at infinity.
Formally it can be treated as a matrix element
\be
D_d(x,x')=\langle x|\hat{D}|x'\rangle
\ee
of the operator
\be\begin{split}\label{operators}
&\hat{D}=-\hat{F}^{-1} \hh \hat{D}=\tilde{D}(-\lap) ,\\
&\tilde{D}(\xi)=-{1\over\tilde{F}(\xi)}={1\over\xi a(-\xi)} .
\end{split}\ee
The momentum space calculations of $D_d(x,x')$ are presented in Appendix
\ref{appB}. The result reads \eq{appDxx}
\be\begin{split}\label{Dxx}
D_d(x,x')={1\over 4\pi}\int_0^\infty
d\eta\,&\tilde{D}(\eta)\,\left({\sqrt{\eta}\over
2\pi |x-x'|}\right)^{{d\over 2}-1}\\
&\times J_{{d\over 2}-1}(\sqrt{\eta}|x-x'|) ,
\end{split}\ee

In Sec. \ref{sec5} and Sec. \ref{sec6} we will use this Green function to
study a gravitational field created by ultrarelativistic particles. For this purpose
it is useful to have another representation of the Green function, where the Bessel
function is replaced by its integral representation
\begin{gather*}
J_{\nu}(z)=\left({z\over 2}\right)^{\nu}{1\over 2\pi
i}\int_{c-i\infty}^{c+i\infty}dt\,
t^{-\nu-1}\exp\left(t-{z^2\over 4t}\right),\\
c>0 .
\end{gather*}
Then after the change of the integration variable
\be
t=i\eta\tau \hh \eta>0 ,
\ee
the Green function can be written in the form
\be\begin{split}\label{Dd}
D_d(x,x')={1\over 2\pi}&\int_0^\infty
d\eta\,\tilde{D}(\eta)\\
&\times\int_{-\infty-ic}^{\infty-ic} {d\tau\over(4\pi
i\tau)^{d/2}}e^{i\tau\eta+i{(x-x')^2\over 4\tau}}.
\end{split}\ee
Note that the last integral contains the expression which is known as the heat
kernel of the Laplace operator in a $d$-dimensional flat Euclidean space
\be
K_d(x,x'|\tau)={1\over(4\pi i\tau)^{d/2}}e^{i{(x-x')^2\over 4\tau}}.
\ee
The heat kernel obeys the equation
\be
i\partial_{\tau}K_d(x,x'|\tau)+\lap K_d(x,x'|\tau)=0
\ee
and the condition
\be
\lim_{\tau\to 0}K_d(x,x'|\tau)=\delta^d(x-x') .
\ee
It describes the amplitude
\be
K_d(x,x'|\tau)=\langle x|e^{i\tau\lap}|x'\rangle .
\ee

In flat space, because of the symmetries of the system in question, both the Green
function $D_d$ and the potential $\psi_d$ are the functions of a distance $r$ between
the points only
\be\label{Ddr}
D_d=D_d(r)\hh \psi_d=\psi_d(r)\hh r=\sqrt{(x-x')^2} .
\ee
The potential at the point $x$ created by the massive particle located at the point
$x'$ is
\be\label{psid0}
\psi_d=-\kappa_d  m\,D_d(r) .
\ee


\section{Gravitational potential in linearized GF gravity theories}\label{sec4}

\subsection{General properties of GF theories}

All GF theories of gravity are assumed to reproduce Einstein gravity in the low
energy regime, i.e., at large scales. In particular it means that the functions
$a(\xi)$ and $c(\xi)$ approach smoothly to 1 at small $\xi$:
\be
a(\xi)=1+O(\xi)\hh  c(\xi)=1+O(\xi) .
\ee
Then we have the functions $\tilde{F}(\xi)=-\xi+O(\xi)$ and 
$\tilde{D}(\xi)=1/\xi+O(1)$. This
property and \eq{Dxx},\eq{psid0} guarantee that in the limit of large distances one
gets a universal asymptotic for the potential for all these GF theories:
\be\label{psilarge}
\psi_d(r)\Big|_{r\rightarrow\infty}=-\kappa_d  m\,{\Gamma\left({d\over
2}-1\right)\over 4\pi^{d/2}r^{d-2}} .
\ee
Obviously, as it should be, it exactly reproduces the gravitational potential
\eq{NGP} in the higher-dimensional Einstein gravity theory.

The asymptotic of the potential at small distances is theory dependent.
Our particular interest is in
$\mathrm{GF_\inds{N}}$ theories, where
\be\label{axi}
a(-\xi)=\exp((\xi/\mu^2)^N)
\ee
and $N=1$ or an even integer number.
The parameter $\mu$ characterizes the scale where the nonlocality becomes important.
One can show that for all
$\mathrm{GF_\inds{N}}$ gravities the potential $\psi_d$ is finite at small $r$. For 
these theories
the asymptotic at $r\mu\rightarrow 0$ can be computed explicitly.
Let us substitute \eq{axi} to \eq{operators},\eq{Dxx},\eq{psid0} and change the
integration variable $\eta={z^2/r^2}$. Then we have
\be\label{psiN}
\psi_d(r)=-{\kappa_d m\over (2\pi)^{d/2}r^{d-2}}\int_0^\infty dz\, z^{{d\over
2}-2} e^{-{z^{2N}\over r^{2N}\mu^{2N}}}J_{{d\over 2}-1}(z) .
\ee

One can see that in the limit when $r\mu\rightarrow 0$ only small arguments of the
Bessel function contribute to the integral \eq{psiN}. Therefore, one can substitute
there an expansion
\be
J_{{d\over2}-1}(z)={\left({z\over 2}\right)^{{d\over
2}-1}\over\Gamma(d/2)}\left[1-{z^2\over 2d}+{z^4\over 8
d(d+2)}+O(z^6)\right] .
\ee
Then taking the integrals in \eq{psiN} one obtains
\be\label{psiassympt}
\psi_d(r)\sim-\kappa_d  m{\mu^{d-2}\left[\Gamma\left({d-2\over
2N}\right)-{r^2\mu^2\over 2d}\Gamma\left({d\over 2N}\right)\right]\over
(4\pi)^{d/2}N
\,\Gamma\left({d\over 2}\right)}+O(r^4\mu^4) .
\ee
One can see that the leading term is finite and proportional to $\kappa_d
m\mu^{d-2}$. Moreover
the next term in the expansion is proportional to $r^2$ that guarantees regularity of
the metric at $r=0$.

There are other interesting universal properties of the potentials in generic GF
gravities. For example, because the distance $r$ in the integral \eq{Dxx} does not
enter the function $\tilde{D}$ and due to the properties of the derivatives of Bessel
functions it is clear that there is a universal relation
\be\label{d2d}
D_{d+2}(r)=-{1\over 2\pi r} {\partial\over\partial r}D_d(r) .
\ee
For the potentials, considered as functions of the radial distance $r$, this property 
leads to the relation
\be\label{psi2psi}
{1\over \kappa_{d+2}}\psi_{d+2}(r)=-{1\over \kappa_{d}}{1\over 2\pi r}
{\partial\over\partial r}\psi_d(r) ,
\ee
provided the mass parameter $m$ is the same in $d$ and $(d+2)$ dimensions.


\subsection{Potential in $\mathrm{GF_1}$ theory}

The static potential $\psi_d$ in the $\mathrm{GF_1}$ theory satisfies the equation
\be
\exp(-\lap/\mu^2)\lap\psi_d=\kappa_d  m\delta^d(x-x') ,~
\ee
so that
\be
\tilde{F}(\xi)=-\xi e^{\xi/\mu^2}\hh
\tilde{D}(\xi)={1\over\xi}e^{-\xi/\mu^2} .
\ee
Substitution of this expression into \eq{Dxx} and change of the integration variable
$\eta=z^2/r^2$
leads to
\be\begin{split}
D_d(r)&={1\over (2\pi)^{d/2}r^{d-2}}\int_0^\infty dz\,
z^{{d\over 2}-2} e^{-{z^2\over r^2\mu^2}}J_{{d\over
2}-1}(z)\\
&={\gamma\left({d\over
2}-1,{r^2\mu^2\over 4}\right)\over 4\pi^{d/2}r^{d-2}} ,
\end{split}\ee
where $\gamma(n,x)$ is the lower incomplete gamma function \cite{Olver:2010}. At
large distance
$r\gg\mu^{-1}$ this expression reproduces the static Green function of the
$d$-dimensional Laplace operator
\be
G_d(x,x')={\Gamma\left({d\over 2}-1\right)\over 4\pi^{d/2}r^{d-2}} .
\ee
For small distances $r\ll\mu^{-1}$ the Green function $D_d(r)$ is a regular function 
of $r$ and is of the form
\be
D_d(r)={2\mu^{d-2}\over (d-2)(4\pi)^{d/2}}\left(1-{d-2\over d}r^2\mu^2\right)+\dots
~ .
\ee

The potential $\psi_d$ is given by
\be
\psi_d=-\kappa_d  m\,D_d(x,x')=-\kappa_d  m\,
{\gamma\left({d\over 2}-1,{r^2\mu^2\over 4}\right)\over 4\pi^{d/2}r^{d-2}} .
\ee
In four-dimensional spacetime ($d=3$) we reproduce the results of
\cite{Biswas:2011ar,Modesto:2010uh,Frolov:2015bia,Nicolini:2005zi,Gruppuso:2005yw}
\be
\psi_3=-\kappa_3  m\,{\mathrm{erf}\left({r\mu/ 2}\right)\over 4\pi r} .
\ee
In the case of five-dimensional spacetime ($d=4$)  we obtain even simpler
expression
\be
\psi_4=-\kappa_4  m\,{1-\exp\left(-r^2\mu^2/4\right)\over 4\pi^2 r^2} .
\ee

The potentials $\psi_d$ in an arbitrary number of dimensions qualitatively look
alike. They are negative
and finite at $r=0$. At larger distances they become more shallow and at
$r\gg\mu^{-1}$ quickly approach the Einstein asymptotic \eq{psilarge}.


\subsection{Potential in $\mathrm{GF_2}$ theory}

When $N=2$ the operator $\hat{F}$ corresponds to
\be
a({\lap})= \exp(\lap^2/\mu^{4})
\ee
and, hence,
\be
\tilde{F}(\xi)=-\xi e^{\xi^2/\mu^4}\hh
\tilde{D}(\xi)={1\over\xi}e^{-\xi^2/\mu^4} .
\ee
Then the potential takes the form
\be\begin{split}
\psi_d(r)&=-{\kappa_d  m\,\mu^{d-2}\over d(d-2)\,2^{{3d\over 2}-2}\pi^{d-1\over 2}} 
\\
&\times\left[{d\over\Gamma\left({d\over
4}\right)}{}_\ins{1\!}F_\ins{3}\left({d\over
4}-{1\over 2};{1\over 2},{d\over 4},{d\over 4}+{1\over 2};y^2\right)
\right.\\
&\left.-{2(d-2)\,y\over\Gamma\left({d\over 4}+{1\over
2}\right)}{}_\ins{1\!}F_\ins{3}\left({d\over 4};{3\over 2},{d\over 4}+1,{d\over
4}+{1\over 2};y^2\right)
\right] ,
\end{split}\ee
where
\be
y={r^2\mu^2\over 16} .
\ee
and ${}_\ins{p\!}F_\ins{q}$ is the generalized hypergeometric function
(see, e.g.,\cite{Olver:2010}).

\begin{figure}[tbp]
\centering
\includegraphics[width=8cm]{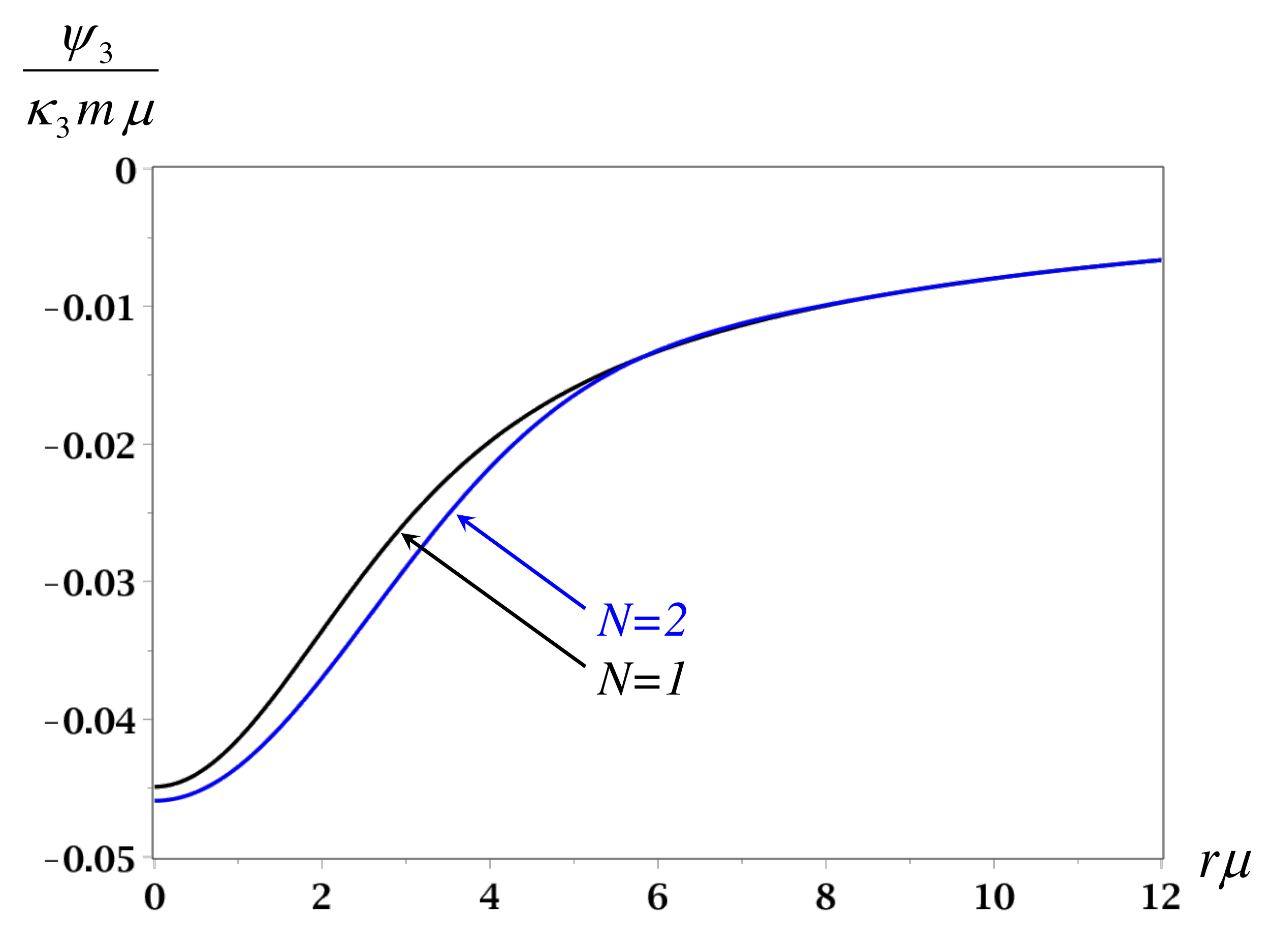}
  \caption{This plot shows the function $\psi_3(r)$ for $N=1$ and $N=2$.
\label{psi3a}}
\end{figure}

\begin{figure}[tbp]
\centering
\includegraphics[width=8cm]{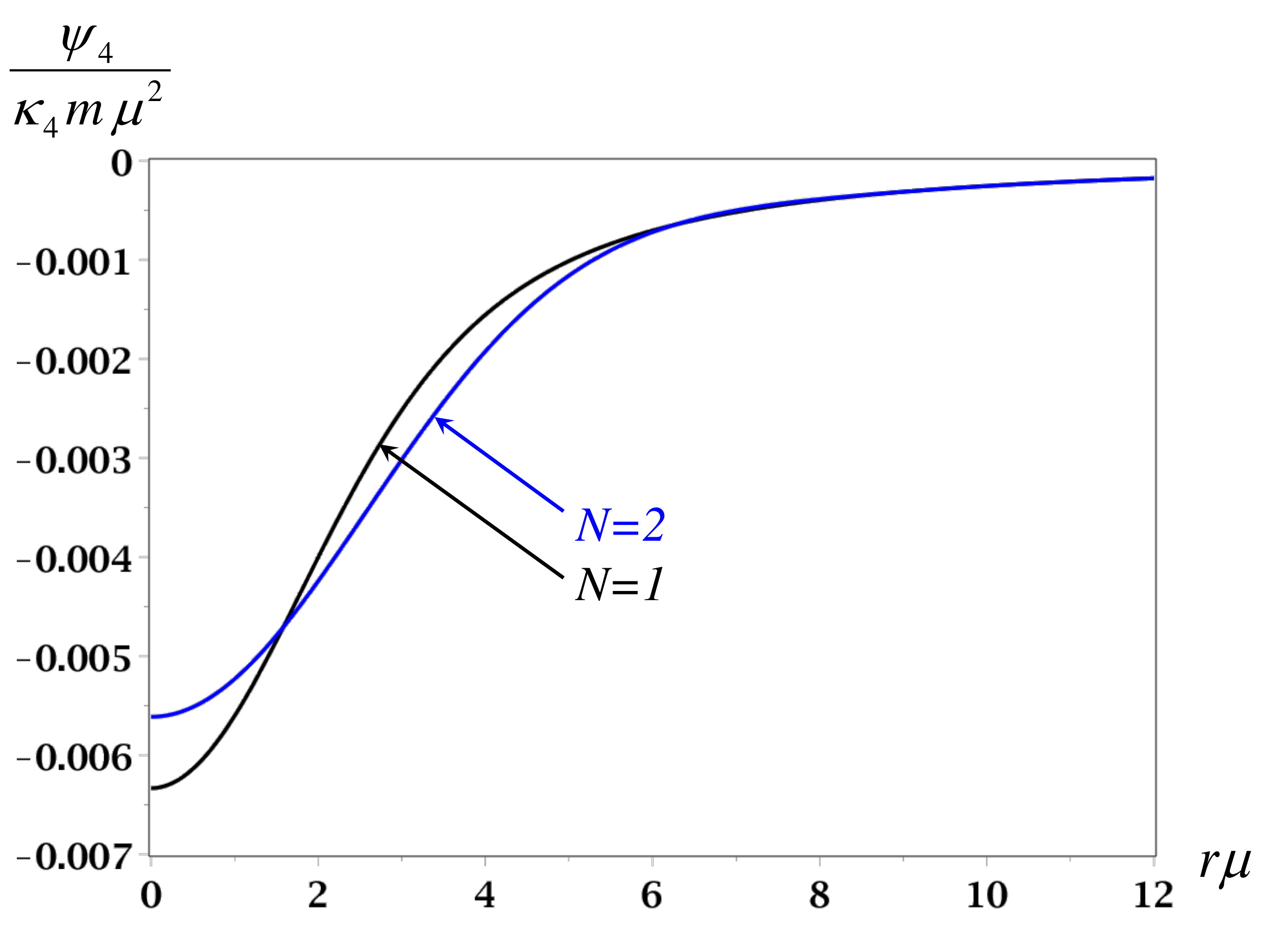}
  \caption{
  This plot shows the function $\psi_4(r)$ for $N=1$ and $N=2$.
\label{psi3b}}
\end{figure}

Qualitatively the potentials for different parameters $N$ and in different
dimensions
$d$ look similar. Figs.~\ref{psi3a}-\ref{psi3b} show examples of the
gravitational potential for
$d=3$ and $d=4$ in two cases, $N=1$ and $N=2$.

\subsection{Potential in $\mathrm{GF_\inds{N}}$ theories}

Similar results in terms of the generalized hypergeometric functions can be derived
for an arbitrary $\mathrm{GF_\inds{N}}$ theory. For all these theories the asymptotic
at large distances is governed by the \eq{psilarge} and the asymptotic at small
distances is given by \eq{psiassympt}. Let us present here only one more explicit
example of the potential in $\mathrm{GF_4}$ gravity
\be\begin{split}
\psi_d(r)=-A\left[d(d+2)(d+4)^2\Gamma\left({d-2\over 8}\right)B_1\right.~&\\
\left.
-8y(d+2)(d+4)^2\Gamma\left({d\over 8}\right)B_2\right.~&\\
\left.
+32y^2(d+4)^2\Gamma\left({d+2\over 8}\right)B_3\right.~&\\
\left.
-{2048\over 3}y^3\Gamma\left({d+12\over 8}\right)B_4
\right]& ,
\end{split}\ee
where
\begin{gather*}
B_1={}_\ins{1\!}F_\ins{7}\scriptstyle\left({d-2\over 8};{1\over 4},{1\over
2},{3\over 4},{d\over
8},
{d\over 8}+{3\over 4},{d\over 8}+{1\over 2},{d\over 8}+{1\over
4};{y^4\over 256}\right),\\
B_2={}_\ins{1\!}F_\ins{7}\scriptstyle\left({d\over 8};{1\over 2},{3\over
4},{5\over 4},{d\over
8}+1,
{d\over 8}+{3\over 4},{d\over 8}+{1\over 2},{d\over 8}+{1\over 4};{y^4\over
256}\right),\\
B_3={}_\ins{1\!}F_\ins{7}\scriptstyle\left({d+2\over 8};{3\over 4},{5\over
4},{3\over 2},{d\over
8}+1,
{d\over 8}+{3\over 4},{d\over 8}+{1\over 2},{d\over 8}+{5\over 4};{y^4\over
256}\right),\\
B_4={}_\ins{1\!}F_\ins{7}\scriptstyle\left({d+4\over 8};{5\over 4},{3\over
2},{7\over 4},{d\over
8}+1,
{d\over 8}+{3\over 4},{d\over 8}+{3\over 2},{d\over 8}+{5\over 4};{y^4\over
256}\right),
\end{gather*}
and the coefficient
\begin{gather*}
A={\kappa_d  m\,\mu^{d-2}\over 2^{2d-{1\over 2}}\pi^{d-3\over
2}\,d(d-2)(d+2)(d+4)}\\
\times{1\over
\Gamma\left({d-2\over 8}\right)\Gamma\left({d\over 8}\right)
\Gamma\left({d+2\over 8}\right)\Gamma\left({d+12\over 8}\right)}  .
\end{gather*}
Expressions for the potentials become more complicated for higher $N$ and we do not
present them here.

\section{Penrose limit}\label{sec5}

Let us demonstrate now, that obtained static solutions of the GF
gravity can be used to find the gravitational field of an ultrarelativistic object.
In the standard 4D Einstein gravity such a limiting metric is known as an
Aichelburg-Sexl metric \cite{Aichelburg:1970dh}. This metric was generalized to the 
case of higher
dimensions and for the spinning objects (called gyratons)
in papers \cite{Frolov:2005in,Frolov:2005zq,FrolovZelnikov:2011}.
In this section we obtain a metric created by an ultrarelativistic object
moving in $D$-dimensional spacetime (nonspinning gyraton metric) in $GF$ theories
of
gravity. As we shall see a key role in this derivation is played by the heat kernel
representation \eq{Dd} of the Green function $D_d(x,x')$.

Consider the metric in the following form
\be\begin{split}\label{metric1}
ds^2=-&(1+2\varphi_d)\,dt^2+(1-2\psi_d+2\varphi_d)(dy^2+d\zeta^2_\perp) , \\
&x=(y,\zeta_\perp)\hh \zeta_\perp=(\zeta^2,\dots,\zeta^{d+1}) .
\end{split}\ee
Let us boost this metric in the $y$-direction
\be
\bar{t}=\gamma(y-\beta t)\hhh \bar{y}=\gamma(t-\beta y)\hhh
\gamma=(1-\beta^2)^{-1/2} ,
\ee
and introduce null coordinates
\be
u=\bar{t}-\bar{y}\hh v=\bar{t}+\bar{y} .
\ee
In the relativistic limit, when the
boost velocity is close to the speed of light, i.e.,  $\beta\rightarrow 1$, the
boost factor $\gamma\rightarrow\infty$. In this limit
$dt\sim\gamma du$ and $dy\sim-\gamma du$. Then the line element \eq{metric1} becomes
\be\label{metric2}
ds^2=-dudv+d\zeta^2_\perp+\varPhi_d du^2 ,
\ee
where
\be
\varPhi_d=-2\lim_{\gamma\rightarrow \infty}(\gamma^2\psi_d) .
\ee
For a point particle of mass $m$ the
Penrose limit corresponds to ultrarelativistic limit $\gamma\rightarrow\infty$ with
the condition that an energy $E=\gamma m$ of the particle is kept fixed.

The gravitational potential $\psi_d$ (see \eq{psid0},\eq{Dd}) can be presented in
the form
\be\begin{split}\label{psid1}
\psi_d&=-\kappa_d  m D_d(r)\\
&=-{\kappa_d  m\over 2\pi}\int_0^{\infty}d\eta\,
\tilde{D}(\eta)\int_{-\infty}^{\infty}{d\tau\over (4\pi
i\tau)^{d/2}}e^{i\eta\tau}\,e^{i{r^2\over 4\tau}} .
\end{split}\ee
One can see that the boost affects only the last exponent in this integral 
representation.

Taking into account that after the boost
\be
y\rightarrow-\gamma u\hhh
r^2\rightarrow \gamma^2 (u-u')^2+\rho^2\hhh
\rho^2=(\zeta_\perp-\zeta_\perp')^2 ,
\ee
and using the delta-function representation
\be
\lim_{\gamma\rightarrow \infty}{\gamma\over\sqrt{4\pi i\tau}}e^{i{\gamma^2u^2\over
4\tau}}=\delta(u) ,
\ee
we obtain
\be
\varPhi_d={\cal F}_d(\rho)\delta(u-u') .
\ee
Here
\be
{\cal F}_d(\rho)={\kappa_d  E\over \pi}\int_0^{\infty}d\eta\,
\tilde{D}(\eta)\int_{-\infty}^{\infty}{d\tau\over (4\pi
i\tau)^{(d-1)/2}}e^{i\eta\tau}\,e^{i{\rho^2\over 4\tau}} .
\ee
Comparison of this integral expression with \eq{psid1} leads to the observation 
that the function
${\cal F}_d(\rho)$ is proportional to the gravitational potential defined in space of
one dimension less, i.e., in the space orthogonal to the particle motion:
\be
{\cal F}_d(\rho)=2\kappa_d  E\,D_{d-1}(\rho)=-2{\kappa_d E\over
\kappa_{d-1}m}\psi_{d-1}(\rho) .
\ee
This property is valid for arbitrary $\mathrm{GF_\inds{N}}$  theories of gravity.

Using the property \eq{psi2psi}, which is also valid for a generic 
$\mathrm{GF_\inds{N}}$ gravity,
we derive a relation
\be\label{aa3}
\partial_\rho{\cal F}_d(\rho)=4\pi\tilde{E}\,\rho\,\psi_{d+1}(\rho)\hh
\tilde{E}={\kappa_d E\over \kappa_{d+1}m} .
\ee
This relation will be useful for the study of gravitational effects in collisions of
ultrarelativistic particles (gyratons
\cite{Frolov:2005zq,Yoshino:2007ph,FrolovZelnikov:2011}) in the next sections.

\section{Apparent horizon formation for head-on collision of the ultrarelativistic
particles}\label{sec6}

\begin{figure}[tbp]
\centering
\includegraphics[width=8cm]{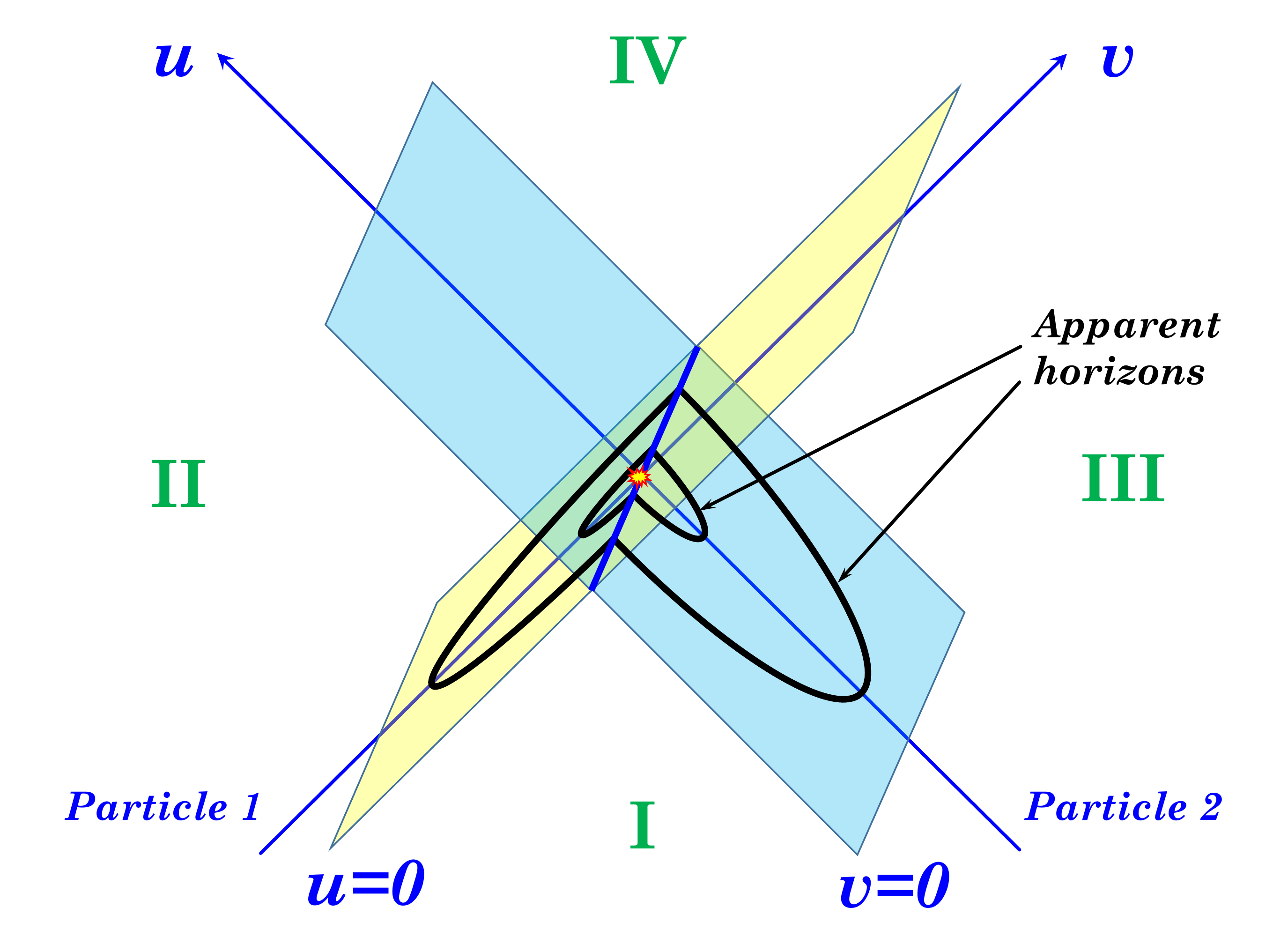}\hfill
  \caption{Head-on collision of two ultrarelativistic particles.
\label{Fig_1}}
\end{figure}

Our next goal is to use the obtained results to study head-on collision of the
ultrarelativistic particles in the GF theories of gravity. We use an approach
developed by Penrose \cite{Penrose} and D'Eath and Payne
\cite{D'Eath:1992hb,D'Eath:1992hd,D'Eath:1992qu} and approximate the colliding
particles by {\em gyratons}. A schematic picture of such a process is shown in
Fig.~\ref{Fig_1}. It shows two-particle motion in the center-of-mass frame. Each of
the particles moves with the velocity of light. Particle 1 moves from the left to the
right along the $y$-direction, while particle 2 moves in the opposite direction. The 
null
lines, representing their trajectories, belong to $u=0$ and $v=0$ null planes,
correspondingly. The gravitational field of these particles is localized on the plane
$u=0$ (for particle 1) and $v=0$ (for particle 2). The intersection of two null
planes is the $(d-1)$-dimensional transverse plane. In the regions $I$, $II$, and
$III$,
outside the $u=0$ and $v=0$ null planes the metric is flat and null rays in these
domains are nothing but null straight lines. However, when such a ray passes either
through  $u=0$ or $v=0$ planes, it is scattered by the gravitational field of the
corresponding particle.

Our purpose is to study formation of the apparent horizon in such a process. Let us
remember that a {\em trapped surface} is a compact spacelike $(d-1)$-dimensional
surface which has the property that both of the null congruences orthogonal to it,
are
not expanding. We focus on the outgoing congruence. One calls a trapped surface a
{\em marginally trapped surface} if the outer normals to it have zero convergence
\cite{Hawking:1973uf}. In a spherically symmetric spacetime one may consider
spherical slices and define an apparent horizon as $d$-dimensional surface which on
each of the slices coincides with the marginally trapped surface.

The problem of ultrarelativistic particle collision in general relativity was
discussed recently in connection with possible mini-black-hole creation in colliders
\cite{Eardley:2002re,Yoshino:2002tx,Yoshino:2005hi,Yoshino:2007ph}.  Eardley and
Giddings
\cite{Eardley:2002re} demonstrated that a problem of existence of the apparent
horizon can be reduced to a special boundary-value problem for an elliptic (Poisson)
equation in a flat spacetime.  Generalizations of
these results to the collision of shock waves on AdS background were
also considered in \cite{Gubser:2008pc,Kiritsis:2011yn}.  The problem is
greatly simplified for the case of the
head-on collision and can be solved analytically in any number of spacetime
dimensions. In the present paper we follow their approach. Let us write the metric
(\ref{metric2}) in the form \ba\n{Yoshino:2005hi}
ds^2&=&-d\bar{u}d\bar{v}+d\bar{\zeta}^2_\perp+\varPhi_d d\bar{u}^2\, ,\\
\varPhi&=&{\cal F}_d(\bar{\rho})\delta(\bar{u})\hh 
\bar{\rho}=\sqrt{\bar{\zeta}_i^2}\, .
\ea
It is possible to show that geodesics and their tangent vectors are not continuous in
these coordinates (see e.g. \cite{Eardley:2002re}). One can change the coordinates so
that both geodesics and their tangent vectors will be continuous in the new
coordinates. The new coordinates in the domain $II$ are defined as follows
\be\begin{split}
\bar{u}&=u \hh \bar{\zeta}_i={\zeta}_i+{u\over 2}\nabla_i\varPhi \, \vartheta(u)\,
,\\
\bar{v}&=v+\varPhi \, \vartheta(u) +{1\over 4} u\, \vartheta(u) \,
(\nabla\varPhi)^2\, .
\end{split}\ee
A similar transformation (with a change $u\leftrightarrow v$) should be made in the 
domain $III$.

The metric (\ref{Yoshino:2005hi}) in the new coordinates takes the form
\be\begin{split}
&ds^2=-du \,dv
+[H_{ik}^{(1)}H_{jk}^{(1)}+H_{ik}^{(2)}H_{jk}^{(2)}-\delta_{ij}]d{\zeta}_i
d{\zeta}_k ,\\
&H_{ij}^{(1)}=\delta_{ij}+{1\over 2}\nabla_i\nabla_j \varPhi \, u\,
\vartheta(u),\\
&H_{ij}^{(2)}=\delta_{ij}+{1\over 2}\nabla_i\nabla_j \varPhi \, v\,
\vartheta(v)\, .
\end{split}\ee

We consider a special marginally trapped surface ${\cal S}$ which consists of two
parts ${\cal S}_u$ ${\cal S}_v$. In coordinates $(u,v,{\zeta}_i)$ a position of
${\cal S}_u$ and ${\cal S}_v$ on two incoming null planes is described by equations
\be\n{eqq}
\{v=-\Psi(\rho), u=0\} \mbox{\ \ \ and\ \ \ } \{u=-\Psi(\rho), v=0\}\, ,
\ee
respectively. These two $(d-1)$-dimensional surfaces intersect at $(d-2)$-dimensional
boundary ${\cal C}$, located at $u=v=0$. The function $\Psi$ is positive inside the
boundary ${\cal C}$ and vanishes at ${\cal C}$. The internal (induced) geometry of
${\cal S}_u$ and ${\cal S}_v$ are the geometry of a half of a $(d-1)$-dimensional
round sphere, their intersection ${\cal C}$ being a round $(d-2)$-dimensional
sphere. For the head-on collision the function $\Psi(\rho)$, which enters both
equations in (\ref{eqq}), is the same. In \cite{Eardley:2002re} it was shown that the
outer null normals have zero convergence in ${\cal S}_u$ and ${\cal S}_v$ if
\be
\nabla^2 (\Psi-{\cal F}_d)=0\, .
\ee
A condition that both normals (in ${\cal S}_u$ and ${\cal S}_v$) coincide at their
boundary ${\cal C}$ implies \be\n{a44}
(\nabla\Psi)^2=4\, .
\ee

\begin{figure}[tbp]
\centering
\includegraphics[width=8cm]{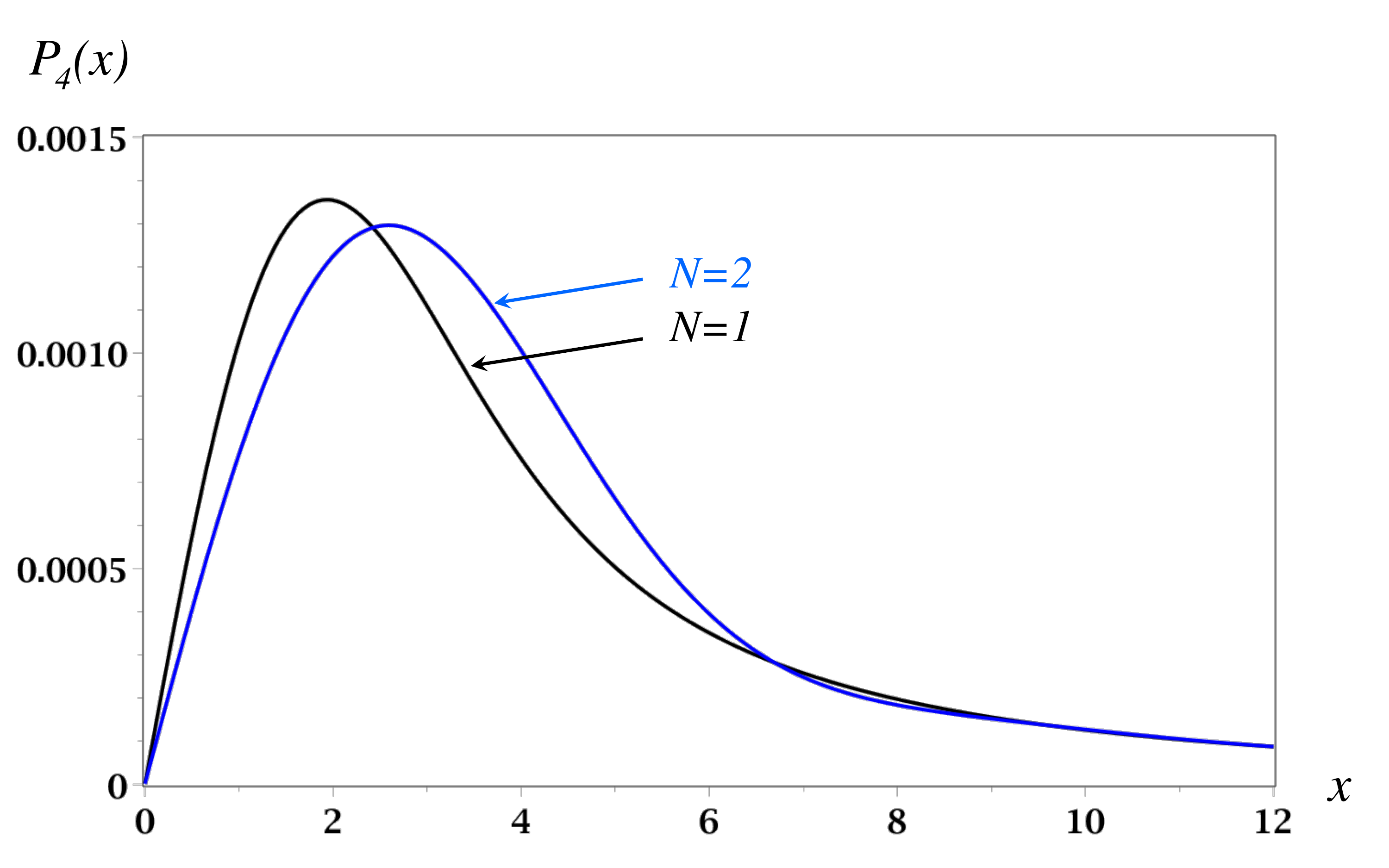}
\caption{The plot shows function $P_4(x)$ for N=1,2.
\label{Fig_3a}}
\end{figure}

\begin{figure}[tbp]
\centering
\includegraphics[width=8cm]{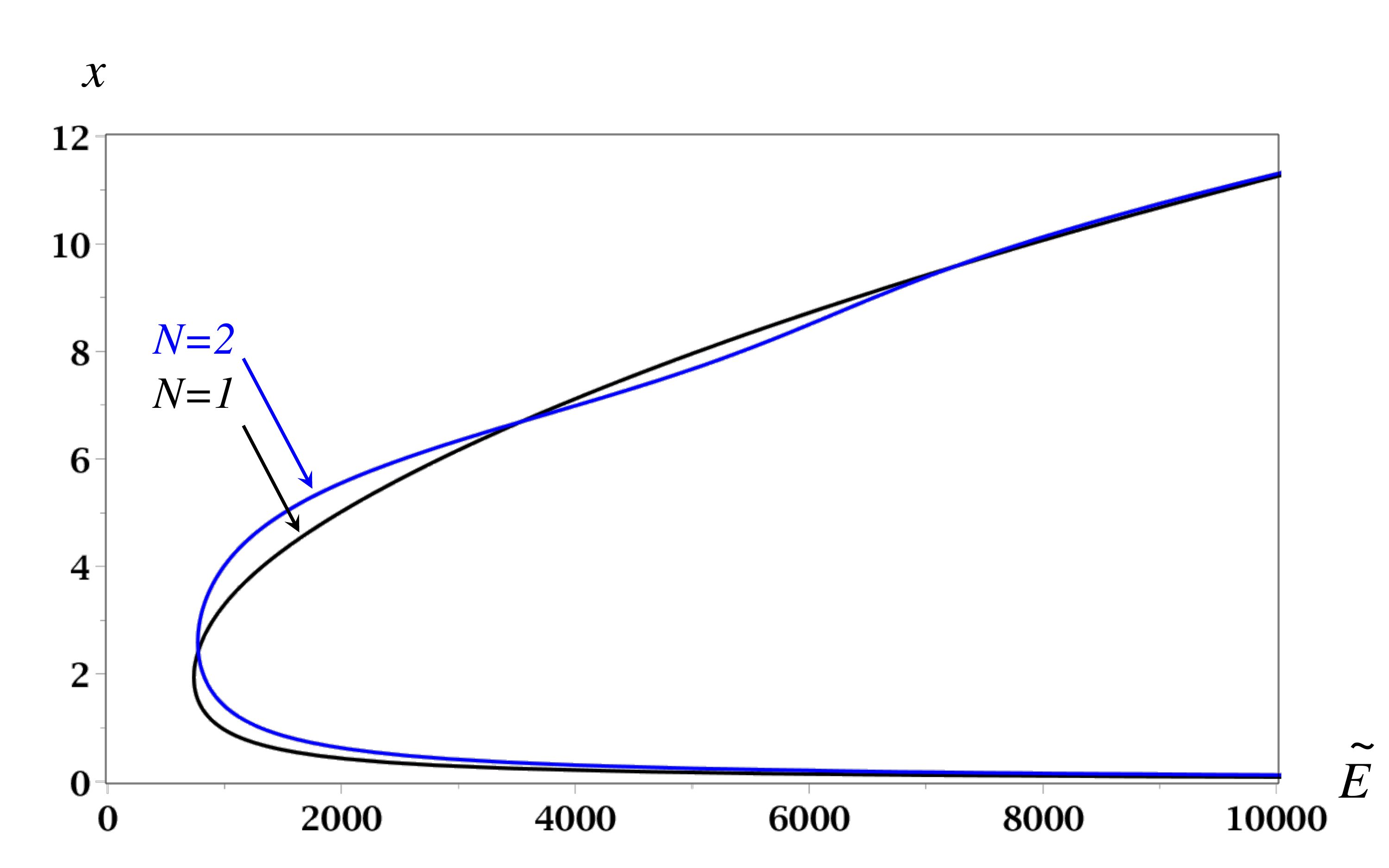}
\caption{The plot shows
the radius
$x=\mu\rho$ of outer and inner apparent horizons as a function of the
energy $\tilde{E}$ for $d=4$ and N=1,2.
\label{Fig_3b}}
\end{figure}

Denote $\chi=\Psi-{\cal F}_d$ and by $\rho_C$ the radius $\rho$ at the boundary.  
Then
\be
\nabla^2 \chi=0\hh \chi_C=-{\cal F}_d(\rho_C)\, .
\ee
Hence one can put $\chi=-{\cal F}_d(\rho_C)$ inside ${\cal C}$ so that
\be
\Psi={\cal F}_d(\rho)-{\cal F}_d(\rho_C)\, .
\ee
The condition \eq{a44} takes the form
\be
\left. (\nabla{\cal F}_d)^2\right|_C=4\, .
\ee
Using \eq{aa3} one gets
\be
2\pi \kappa_d E\, \rho D_{d+1}(\rho)=1\, .
\ee
In terms of a dimensionless coordinate
$
x=\mu\rho
$,
dimensionless energy $\tilde{E}=2\pi \mu^{d-2}\kappa_d\, E$, and a dimensionless
profile function
\be
P_d(x)\equiv x D_{d+1}(x/\mu)/\mu^{d-1},
\ee
this condition reads
\be\label{PdE}
P_d(x)={1\over \tilde{E}} .
\ee
\begin{figure}[tbp]
\centering
\includegraphics[width=8cm]{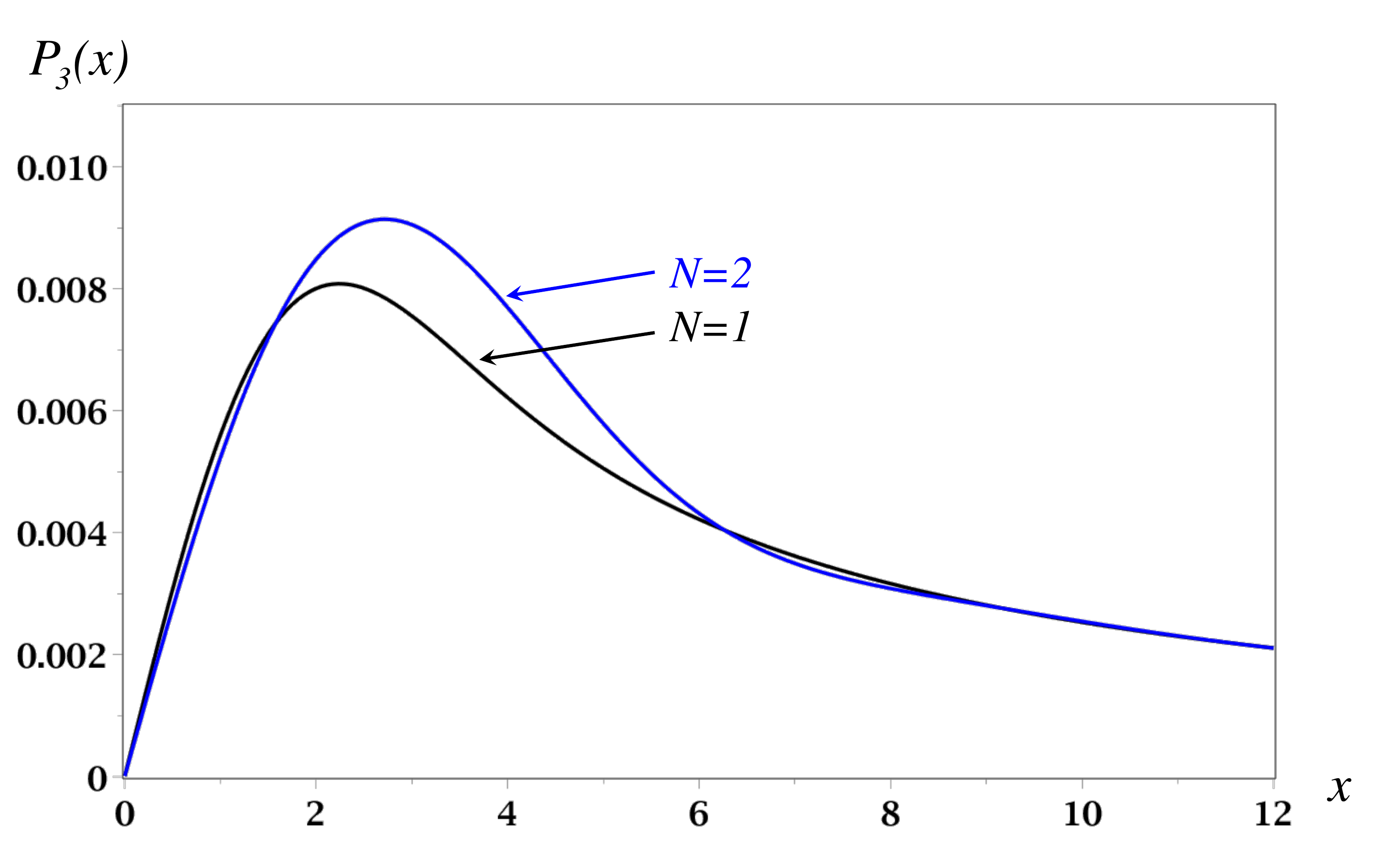}
  \caption{The plot shows function $P_3(x)$ for N=1,2.\label{Fig_2a}}
\end{figure}
\begin{figure}[tbp]
\centering
\includegraphics[width=8cm]{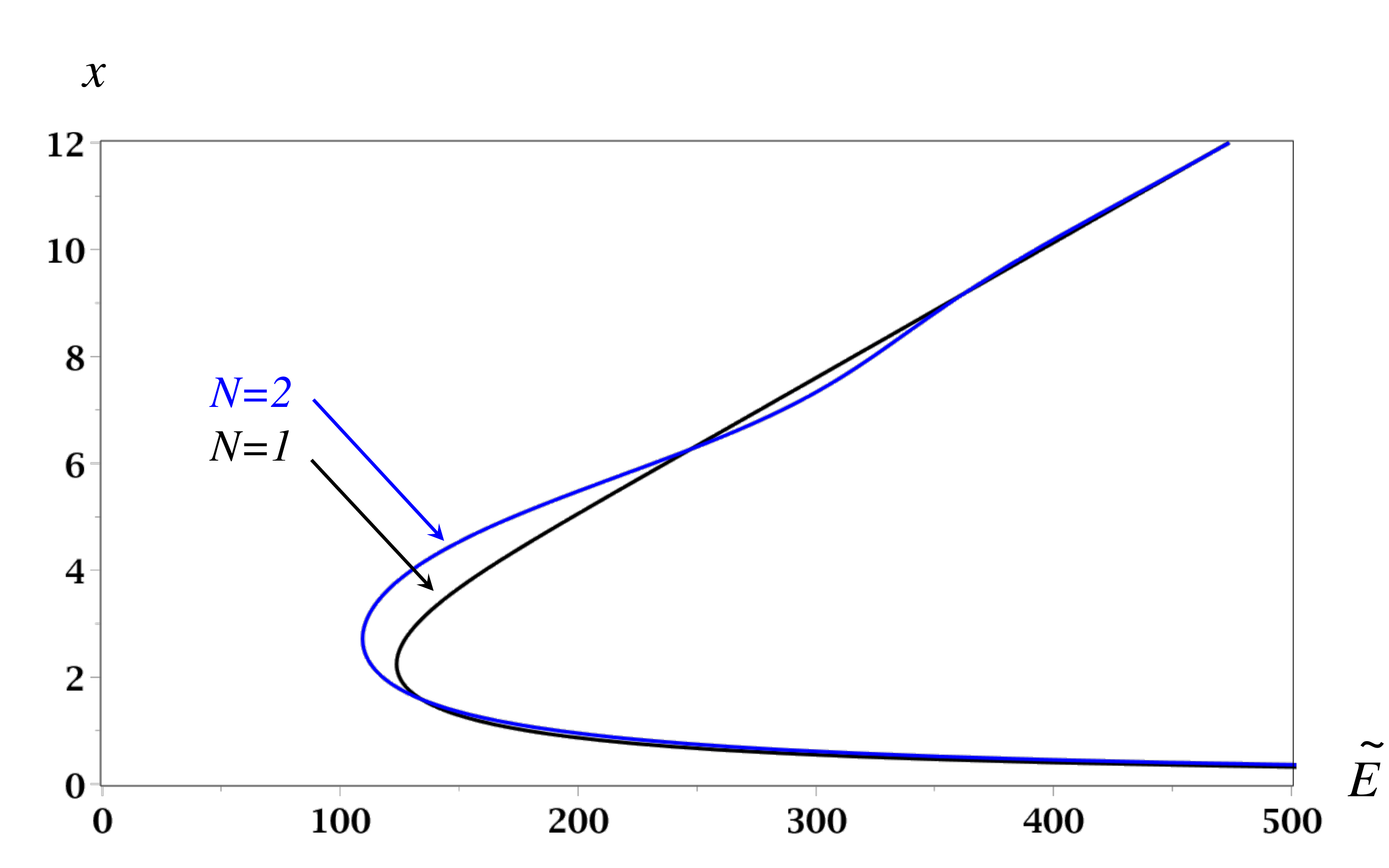}
  \caption{The plot shows
the radius
$x=\mu\rho$ of outer and inner apparent horizons as a function of the
energy $\tilde{E}$ for $d=3$ and N=1,2.}
\label{Fig_2b}
\end{figure}
All functions $P_d(\rho)$ look similar (see
Figs.~\ref{Fig_2a} and \ref{Fig_3a}).
They vanish at $x=0$ and then grow, reach maximum, and then decrease to a universal
asymptotic, that does not depend on
the parameter $N$, though depends on $d$. The plots Figs.~\ref{Fig_2b} and
\ref{Fig_3b}
show solutions of Eq.\eq{PdE}. The apparent horizon
exists for the energy obeying the condition $\tilde{E}\ge
\tilde{E}_\inds{critical}$. In this
energy domain it has at least two branches, inner and outer. At $\tilde{E}=
\tilde{E}_\inds{critical}$
they meet and the apparent horizon disappears \footnote{In the case
of even $N\ge 4$
between inner and outer apparent horizons there may exist  also additional
intermediate apparent horizons.}.
This behavior resembles qualitatively
that of the colliding relativistic extended sources \cite{Taliotis:2012sx}.
This resemblance is not accidental. One can rearrange Laplace opeators in
\eq{Fpsi} and move $a(\lap)^{-1}$ to the
right-hand side of the equation. Then it can be identically rewritten as
\be\label{Fpsi1}
\lap \psi_d=j\hh j=\kappa_d  m\,a(\lap)^{-1}\delta^d(x-x') ,
\ee
When acting on the localized source, the operator $a(\lap)^{-1}$
delocalizes it and makes $j$ to become effectively an extended current for the
traditional Laplace equation \eq{Fpsi1}. In this sense the analogy of effects in
the ghost-free gravities and for the colliding extended sources
\cite{Taliotis:2012sx} becomes evident.


\section{Summary and discussion}\label{summary}

In this paper we discussed an application of the linearized equations of the 
ghost-free theory of gravity to three connected problems. First, we calculated the
gravitational potential of a point mass in the Newtonian limit and showed that GF
modification of gravity works as a regularizer. Namely, this potential is regular
at the origin. This property is valid for $\mathrm{GF_1}$ and 
$\mathrm{GF_{2n}}$ theories in any number
of spatial dimensions $d\ge 3$. This is a generalization of the earlier obtained
result for $\mathrm{GF_1}$ for $d=3$ \cite{Biswas:2011ar,Modesto:2010uh} and for 
$d>3$
\cite{Frolov:2015bia}. The second main result
of the paper is calculation of the gravitational field of an ultrarelativistic
particle in the $\mathrm{GF_\inds{N}}$ theories. The obtained metrics are 
generalizations of the famous
four-dimensional Aichelburg-Sexl metric \cite{Aichelburg:1970dh} of general
relativity. Again, the
obtained metrics are solutions of the equations of the $\mathrm{GF_\inds{N}}$ 
gravity equations
($N=1$ and $N=2n$) in a spacetime with an arbitrary number of  dimensions $d+1$. And
finally, we used these results to study an apparent horizon formation in the head-on
collision of two ultrarelativistic particles. Our main conclusion is that in such a
process there exists a mass gap for the mini-black-hole formation. If $\mu$ is the
characteristic mass scale of the corresponding ghost-free theory, then in order for
a mini-black hole to be formed in the collision, the center-of-mass energy $E$ 
should be
of the order of or larger than $(G^{(d)}\mu^{d-2})^{-1}$. Another important feature 
of
the process is that when the apparent horizon is formed, it has two branches: outer
and inner marginally trapped surfaces. Both of them have the geometry of the sphere.
When the center-of-mass energy increases, the inner part becomes closer to the point
until it reaches the scale $\tilde{\lambda}$, where the model we used breaks down.

This result is again valid for any $\mathrm{GF_\inds{N}}$ theory ($N=1$ and $N=2n$) 
in any number of
dimensions. It can be considered as some indication that for such theories the inner
singularity of a black hole might be absent and there exists a closed apparent
horizon. Such a model was proposed in \cite{Frolov:1981mz} and discussed later
in many publications. It should be emphasized that most of the
results, related to the study of the models with closed apparent horizons,
beyond a linear approximation, were obtained without using concrete dynamical
equations. In this sense they are phenomenological. It is a real challenge to
obtain solutions for a dynamical collapse in the modifications of the Einstein
theory which are UV complete. In particular, in order to
arrive at a definite conclusion concerning the structure of a black hole interior in
the GF gravity one needs to perform analysis in the complete version of such a
theory, which includes nonlinear effects.

\section*{Acknowledgments}

The authors thank the Natural Sciences and Engineering Research Council of Canada and
the Killam Trust for their financial support.

\appendix

\section{Linearized equations of the higher-derivative modification of the
gravitational equations in higher dimensions}\label{appA}

In order to obtain linearized equations of a theory of gravity with higher
derivatives
in higher dimensions one can follow a similar derivation in four dimensions presented
in the papers \cite{Biswas:2011ar,Biswas:2013kla}. In this appendix we collected the
corresponding formulas for further reference.

The main steps of this derivation are the following. One considers first a covariant
action which besides the Einstein term contains also a part $S_q$ which is quadratic
in curvature. The latter may contain an arbitrary number of covariant derivatives 
acting
on each of the curvature tensors. One can always move the derivatives acting on the
first Riemann tensor to the position, when it acts on the other one. This can be
achieved by using integration by parts.
The number of derivatives may even be infinite, so that a theory is nonlocal.
Since each of the Riemann curvature tensors has fourindices, the maximal total 
number of
derivatives with ``free'' indices is eight. All other derivatives can be combined in
functions of the covariant box operator. In order to achieve this it might be
required to commute the derivatives. But this operation produces terms which are of
the third order in the curvature so that they should be neglected in the adopted
approximation. Using symmetry properties of the curvature tensor, Bianchi identities
and commutativity of the covariant derivatives in the adopted approximation one
finally obtains the following expression for $S_q$
\cite{Biswas:2011ar,Biswas:2013kla}
\be\begin{split}\nonumber
S={1\over 2\kappa_d }\int dx\sqrt{-g}&\left[~ R
+RF_1(\Box)R +R_{\mu\nu}F_2(\Box)R^{\mu\nu}\right.\\
&\left.
+R_{\mu\nu\lambda\sigma}F_3(\Box)R^{\mu\nu\lambda\sigma}\right].
\end{split}\ee
Here $\kappa_d =8\pi\,G^\ind{(D)}$ and $G^\ind{(D)}$ is the gravitational coupling
constant in $D$-dimensional spacetime. In four dimensions the value of this constant
is fixed by the requirement that the Poisson equation for the gravitational potential
in the Newtonian limit has a standard form.
There is an ambiguity in the normalization of  $G^\ind{(D)}$ in higher dimensions.
We fix it by requiring the Einstein-Hilbert action to have the same form in all
dimensions.

This general form of the quadratic in curvature action can be further simplified
using the following observation \cite{Barvinsky:1990up,Modesto:2014eta}:
the  "Gauss-Bonnet structures" of the form ($k\ge 1$)
\be\begin{split}
{}^*\!R^{\alpha\beta\gamma\sigma}\,\Box^k\,
{}^*\!R_{\alpha\beta\gamma\sigma}&=R^{\alpha\beta\gamma\sigma}\Box^k
R_{\alpha\beta\gamma\sigma}-4R^{\alpha\beta}\Box^k
R_{\alpha\beta}\\
&+R\,\Box^k R=O(R^3)+\mbox{div}\, .
\end{split}\ee
in arbitrary dimensions are all of the third and higher order in curvature
plus total divergence terms. As a result, the general higher derivative action can
be written in the form which contains only two arbitrary functions of the
box operator \cite{Biswas:2013kla}.

To obtain the linearized equation we write the action in the form
\be\begin{split}\nonumber
S&={1\over 2\kappa_d }\left(S_0+S_1+S_2+S_1+S_3\right)\, ,\\
S_0&=\int dx\sqrt{-g}\,R \, ,\\
S_1&=\int dx\sqrt{-g}\,RF_1(\Box)R\, ,\\
S_2&=\int dx\sqrt{-g}\,R_{\mu\nu}F_2(\Box)R^{\mu\nu}\, ,\\
S_3&=\int
dx\sqrt{-g}\,R_{\mu\nu\lambda\sigma}F_3(\Box)R^{\mu\nu\lambda\sigma}\,.
\end{split}\ee
We use the following expressions for the variations of the objects that enter the
above action and keep only the terms that are quadratic in perturbations
\begin{multline*}
S_0=-\int dx \left(-{1\over 2}h^{\mu\nu}\Box h_{\mu\nu}
+h^{\mu\nu}\partial_{\mu}\partial_{\alpha}\,h^{\alpha}{}_{\nu} \right.\\ \left.
-h^{\mu\nu}\partial_{\mu}\partial_{\nu} h
+{1\over 2}h\Box h\right) ,
\end{multline*}
\begin{multline*}
S_1=\int
dx\left(h^{\mu\nu}\,F_1(\Box)\partial_{\mu}\partial_{\nu}\partial_{\alpha}\partial_{
\beta}\,h^{\alpha\beta}
\right.\\ \left.
-2h^{\mu\nu}\,\Box F_1(\Box)\partial_{\mu}\partial_{\nu}\,h+
h\, \Box^2 F_1(\Box)\, h\right) ,
\end{multline*}
\begin{multline*}
S_2={1\over 4}\int
dx\left(2h^{\mu\nu}\,F_2(\Box)\partial_{\mu}\partial_{\nu}\partial_{\alpha}\partial_{
\beta}\,h^{\alpha\beta}
\right. \\ \left.
-2 h^{\mu\nu}\,\Box F_2(\Box)\partial_{\mu}\partial_{\alpha}\,h^{\alpha}{}_{\nu}
-2h^{\mu\nu}\,\Box F_2(\Box)\partial_{\mu}\partial_{\nu}\,h
\right. \\ \left.
+h^{\mu\nu}\,\Box^2 F_2(\Box)\,h_{\mu\nu}
+h\, \Box^2 F_2(\Box)\, h\right) ,
\end{multline*}
\begin{multline*}
S_3=\int dx\left(
h^{\mu\nu}\,F_3(\Box)\partial_{\mu}\partial_{\nu}\partial_{\alpha}\partial_{\beta}\,
h^{\alpha\beta}
\right. \\ \left.
+h^{\mu\nu}\,\Box^2 F_3(\Box)\,h^{\mu\nu}
-2 h^{\mu\nu}\,\Box F_3(\Box)\partial_{\mu}\partial_{\alpha}\,h^{\alpha}{}_{\nu}
\right) .
\end{multline*}
Let us write the total linearized action $S$ in the form
\be\begin{split}\label{A8}
S={1\over 2\kappa_d }\int dx\left(\phantom{\bigg|}\right.& \!\!
{1\over 2}h^{\mu\nu}\,a\,\Box\,
h_{\mu\nu}+
h^{\mu\nu}\,b\,\partial_{\mu}\partial_{\alpha}\,h^{\alpha}{}_{\nu}\\
&\left. +h^{
\mu\nu }\, c\,
\partial_{\mu}\partial_{\nu} h
+{1\over 2}h\,d\,\Box h
\right.\\
&\left.
+{1\over
2}h^{\mu\nu}\,{f\over\Box}\partial_{\mu}\partial_{\nu}\partial_{\alpha}\partial_{
\beta}\,h^{\alpha\beta}
\right) .
\end{split}\ee
Then we have
\be\begin{split}
a&=1+{1\over 2}F_2\Box+2F_3\Box \,,\\
b&=-1-{1\over 2}F_2\Box-2F_3\Box \,,\\
c&=1-2F_1\Box-{1\over 2}F_2\Box \,, \\
d&=-1+2F_1\Box+{1\over 2}F_2\Box \,, \\
f&=2F_1\Box+F_2\Box+2F_3 \Box \,.
\end{split}\ee
It is easy to see that the form factors $a,b,c,d,f$ satisfy the identities
\be
a+b=0\hh c+d=0\hh b+c+f=0 \,.
\ee

The equations of motion, obtained from \eq{A8}, are
\be\begin{split}\label{eqh}
&a(\Box)\Box
h_{\mu\nu}+b(\Box)\partial_{\sigma}(\partial_{\nu}h_{\mu}{}^{\sigma}+\partial_{\mu}h_
{\nu}{}^{\sigma})\\
&+c(\Box)(\eta_{\mu\nu}\partial_{\rho}\partial_{\sigma}h^{
\rho\sigma}
+\partial_{\mu}\partial_{\nu}h)+\eta_{\mu\nu}d(\Box)\Box
h\\
&+f(\Box)\Box^{-1}\partial_{\mu}\partial_{\nu}\partial_{\rho}\partial_{\sigma}h^
{
\rho\sigma}=-2\kappa_d \tau_{\mu\nu} \,.
\end{split}\ee
Here
\be
\tau^{\mu\nu}={2\over \sqrt{-g}}{\delta S_\ind{Matter}\over\delta g_{\mu\nu}}\, .
\ee
Let us remember that $\eta_{\mu\nu}$ is a metric in $D$-dimensional Minkowski 
spacetime
and partial derivatives and the $\Box$ operator are written in Cartesian coordinates
in this space. Let us emphasize that the number of independent arbitrary functions of
the $\Box$ operator, as well as the form of the equations, is the same as in the
four-dimensional case. However, the dimensional gravitational coupling constant
$G^\ind{(D)}$ depends on the number of dimensions. We also show in Sec. II that the
form of the equations for static gravitational potentials, which contain contractions
of the form $\eta_{\mu\nu}h^{\mu\nu}$, would be explicitly dependent on $D$.


\section{Gravitational potential in momentum space}\label{appB}

Let write the gravitational potential $\psi_d$ in terms of the modes in momentum
space
\be
\psi_d(x)=\int {d^d{\bf k}\over (2\pi)^d}e^{i{\bf k} x}\bar{\psi}({\bf k}).
\ee
Here ${\bf k}=k_i$ is the $d$-dimensional vector of momentum. Similarly we have
\be
D_d(x,x')=\int {d^d{\bf k}\over (2\pi)^d}e^{i{\bf k} (x-x')}\bar{D}({\bf k}).
\ee
From \eq{psid0} and \eq{operators} one can derive
\begin{gather*}
\bar{\psi}({\bf k})=-\kappa_d m \bar{D}({\bf k}),\\
\bar{D}({\bf k})={1\over k^2 a(-k^2)}=\tilde{D}(k^2)\hh k=|{\bf k}|.
\end{gather*}
Using the spherical symmetry of the system we get
\begin{gather*}
d^d{\bf k}=dk d\theta\,k^{d-1}\sin^{d-2}{\theta}\,A_{d-2} ,\\
A_{d-2}=2\,{\pi^{(d-1)/2}\over\Gamma\left({d-1\over 2}\right)},
\end{gather*}
where $A_{d-2}$ is the area of a unit sphere $S^{d-2}$. In spherical coordinates
\be
{k}_i(x^i-x'{}^i)=kr\cos(\theta)\hh r=|{x-x'}|.
\ee
Then the Green function reads
\be\begin{split}\nonumber
D_d(x,x')&=2\,{\pi^{(d-1)/2}\over\Gamma\left({d-1\over 2}\right)}\int_0^{\infty}
{dk\over (2\pi)^d}\,{k^{d-3}\over
a(-k^2)} \\
&\times\int_0^{\pi}
d\theta\,\sin^{d-2}{\theta}\,e^{ikr\cos(\theta)} .
\end{split}\ee
Integration over $\theta$ gives the expression for the Green function $D_d(x,x')$
in terms of an integral from Bessel function
\be
D_d(x,x')={1\over 2\pi}\int_0^\infty {dk\over k a(-k^2)}\left({k\over
2\pi r}\right)^{{d\over 2}-1}J_{{d\over 2}-1}(kr) ,
\ee
Both the potential $\psi_d$ and the Green function $D_d(x,x')$ depend only on the distance $r$ between points.
Change of the integration variables leads to the following equivalent forms ($z=kr$)
\be
D_d(r)={1\over (2\pi)^{d/2}\,r^{d-2}}\int_0^{\infty}
dz\,{z^{d-4\over 2}\over a(-z^2/r^2)}\, J_{{d\over2}-1}(z) ,
\ee
and ($\eta=z^2/r^2$)
\be\label{appDxx}
D_d(r)={1\over 4\pi}\int_0^\infty d\eta\,\tilde{D}(\eta)\,\left({\sqrt{\eta}\over
2\pi r}\right)^{{d\over 2}-1}J_{{d\over 2}-1}(\sqrt{\eta}r) .
\ee



%

\end{document}